\documentclass{IEEEtran}

\ifCLASSINFOpdf
   \usepackage[pdftex]{graphicx}
\else
   \usepackage[dvips]{graphicx}
\fi

\usepackage{multirow}
\usepackage{array}
\usepackage[font=footnotesize]{caption}
\usepackage[cmex10]{amsmath}
\usepackage{bm}
\usepackage{amssymb, amsmath, amsfonts, amsthm}
\usepackage{subcaption}
\usepackage{mathtools}
\usepackage{booktabs}
\usepackage{siunitx}

\ifCLASSOPTIONcompsoc
  \usepackage[caption=false,font=normalsize,labelfont=sf,textfont=sf]{subfig}
\else
  \usepackage[caption=false,font=footnotesize]{subfig}
\fi

\usepackage{float}
\usepackage{verbatim}

\usepackage{acronym}

\newacro{gfm}[GFM]{Grid-Forming}
\newacro{gfl}[GFL]{Grid-Following}
\newacro{gsc}[GSC]{Generalized Swing Control}
\newacro{ibr}[IBR]{Inverter-Based Resource}
\newacro{pll}[PLL]{Phase-Locked Loop}
\newacro{vsm}[VSM]{Virtual Synchronous Machine}
\newacro{voc}[VOC]{Virtual Oscillator Control}
\newacro{dvsm}[dVSM]{Dual Virtual Synchronous Machine}
\newacro{sm}[SM]{Synchronous Machine}
\newacro{ph}[PH]{port-Hamiltonian}
\newacro{cf}[CF]{Complex Frequency}
\newacro{coi}[CoI]{Center of Inertia}
\newacro{dfig}[DFIG]{Doubly-Fed Induction Generator}
\newacro{csig}[CSIG]{Constant Speed Induction Generator}
\newacro{rocof}[RoCoF]{Rate of Change of Frequency}
\newacro{rocov}[RoCoV]{Rate of Change of Voltages}
\newacro{qep}[QEP]{Quadratic Eigenvalue Problem}
\newacro{eac}[EAC]{Equal-Area Criterion}
\newacro{smib}[SMIB]{Single-Machine-Infinite-Bus}
\newacro{bess}[BESS]{Battery Energy Storage Systems}

\newcommand{\bfg}{\boldsymbol}

\begin{document}

\title{Generalized Swing Control Framework for Inverter-based Resources\vspace{-2mm}}

\author{%
  \IEEEauthorblockN{Rodrigo Bernal and Federico Milano\\}%
  \IEEEauthorblockA{School of Electrical \& Electronic Engineering,
    University College Dublin, Ireland \\
    rodrigo.bernal@ucdconnect.ie, federico.milano@ucd.ie}%
  \thanks{This work is supported by Sustainable Energy Authority of Ireland (SEAI) by funding R.~Bernal and F.~Milano under project FRESLIPS, Grant No.~RDD/00681.}%
  \vspace{-3.5mm}
}

\vspace{-5mm}

\maketitle

\begin{abstract}
  This paper proposes a novel control framework designed for \acp{ibr}, denoted as \acf{gsc}.  The proposed \ac{gsc} framework generalizes the definition of \acf{gfm} control schemes and exploits the coupling between active and reactive power dynamics.  To validate the proposed scheme, we conduct extensive time-domain simulations and small-signal analysis using a modified version of the WSCC 9-bus system and a 1479-bus dynamic model of the all-island Irish transmission system.  The case studies focus on evaluating the dynamic performance of the proposed framework under different configurations, including \ac{vsm}, coupled-\ac{vsm} and dual-\ac{vsm} schemes.  To address the nonlinear nature of power system dynamics, sensitivity analysis based on Monte Carlo methods are employed to improve parameter tuning and assess the stability of \ac{gsc} configurations in the studied systems.
\end{abstract}

\begin{IEEEkeywords} 
  Low-inertia power systems, converter-interfaced generation,
  frequency control, swing equation, voltage control.
\end{IEEEkeywords}

\section{Introduction}
\label{sec.introduction}

\subsection{Motivation}
A fundamental difference between \acfp{ibr} and conventional \acfp{sm} is that the dynamic response of \acp{sm} is determined by their electromechanical dynamics, whereas that of \acp{ibr} is governed by their control schemes.  This feature makes \acp{ibr} highly flexible and enables the design of a wide range of control schemes with diverse objectives \cite{found}.  Motivated by the dynamics of \acp{sm} and with the aim of defining a flexible general control of \acp{ibr}, this paper explores the extension of the \ac{sm} swing equation into the complex power domain.

\subsection{Literature Review}

In grids characterized by high $R/X$ ratios or with high penetration of \acp{ibr}, the conventional paradigm of decoupled active power/frequency and reactive power/voltage is not always true.  Reference \cite{DynCoup2024} shows that voltage/frequency stability is impacted by both self- and coupled-dynamics.  Moreover, voltage and frequency oscillate at the same frequency if the voltage-frequency coupled instability occurs.  Recent analytical work has quantified this coupling and its implications for the fast frequency response of \acp{bess} in high-renewable grids \cite{ffr2025}, while new control strategies explicitly address such coupling effects \cite{Laba2022, derfv, dercomp2024, cmplxctrl2025}.

The limitations in flexibility of \acp{sm} and the challenges of weak grids have increased the adoption of \ac{gfm} inverters.  These devices allow control-based energy conversion without relying on physical inertia.  Several \ac{gfm} control strategies have been proposed, including droop-based methods \cite{TOULAROUD20232693,Wei2023}, voltage oscillator control \cite{VOC2022}, and \ac{vsm} among other schemes \cite{VSM_rev2022}.  For the case of \ac{gfl}, \cite{VSClimit2020} shows that there is a limit for this type of converters before the system becomes unstable, and this limit is strongly influenced by the type of grid support the converters provide and their proper tuning.  Moreover, it has been demonstrated in \cite{100gfl2020, bernal2025transientslackcapability} that despite these challenges, \ac{gfl} inverters can support even 100\% inverter-based power systems.

Stability assessment tools of such modern systems are presented in \cite{Trans2023}, which include Lyapunov direct methods, the equal-area criterion, small-signal analysis, data-driven methods, and time-domain simulations, with the latter being the most widely used for highly nonlinear dynamics and the one used within this work.  Even in a simple single-machine infinite-bus model, the transient behavior is more complicated than commonly thought \cite{sun2018equal}.  Thus, advanced  nonlinear control approaches have been proposed such as Lyapunov-based nonlinear control algorithm \cite{BoFan2022} and data-driven methods \cite{datadriven2021}.  Moreover, to provide damping and inertia from \acp{ibr}, new system-level frameworks \cite{DampInert2025} and converter-level controls \cite{GenInert2025} have been developed.  

In \cite{GSE2022}, a generalized swing equation for PLL-based \acp{ibr} is proposed, integrating angle and frequency dynamics into a unified control framework.  In this vein, our work extends the swing equation into the complex domain, enabling simultaneous control of voltage and frequency dynamics.  This formulation leverages the nonlinear power flexibility of \acp{ibr} and introduces an explicit coupling between active and reactive power with voltage magnitude and phase, offering new dynamic features and modeling flexibility.

\subsection{Contributions}
This paper proposes a novel control framework designed for \acp{ibr}, denoted as \acf{gsc}.  Leveraging \ac{ibr} inherent flexibility, \ac{gsc} extends \ac{sm} energy conversion principles through a complex formulation.
Furthermore, the \ac{gsc} framework uses the concept of \emph{coupled inertia, damping} and \emph{stiffness}, which describe the direct coupling between active and reactive power dynamics.
Additionally, we propose a performance index $\mu_{\rm ts}$ that captures the dynamic performance of relevant variables in transient conditions.  

\subsection{Paper Organization}
The remainder of the paper is organized as follows.  The proposed \ac{gsc} is described in Section~\ref{sec_control}.  Case studies are presented in Section~\ref{sec_cases}, where the features of a set of possible \ac{gsc} configurations are studied.  Finally, conclusions and future work are discussed in Section~\ref{sec_conclusion}.

\subsection{Notation}

All variables are assumed to be time dependent unless explicitly stated.  The symbol $'$ is used to refer to the time derivative.  Complex vectors are denoted with $\bar u$, vectors with $\bfg{u}$, and matrices with $\mathbf U$.

\section{Proposed Control}
\label{sec_control}

Let $\bfg{x}\ \in \mathbb{R} ^2$ denote the generalized positions of a second order system defined by the following dynamics:
\begin{align}
    \bfg{\mathrm{M}}{\bfg x}''+\bfg{\mathrm{D}} (\bfg{x}'-\bfg{x}_o')+\bfg{\mathrm{K}}{\bfg (\bfg{x}-\bfg{x}_o)}=\bfg{f} \label{eq_secord},
\end{align}
where $\bfg{\mathrm{M}}$ is known as the inertia matrix, $\bfg{\mathrm{D}}$ the damping matrix, $\bfg{\mathrm{K}}$ is the stiffness matrix, and $\bfg{f}$ accounts for inputs \cite{Quadratic2001}.  
By rewriting the system as a system of first order differential equations, and introducing 2 new state variables, we obtain the following structure:
\begin{align}
    \bfg{\zeta}'= 
    \begin{bmatrix}
        0_{2\times2} & {\bf I}_{2\times2}\\
        -{\bf M}^{-1}{\bf K} & -{\bf M}^{-1}{\bf D}
    \end{bmatrix} \bfg{\zeta}-
    \begin{bmatrix}
        0_{2\times1}\\
        \bfg{f}
    \end{bmatrix}  \label{eq_secord_2}
\end{align}

where $\bfg{\zeta}=\begin{bmatrix}\bfg{x}-\bfg{x}_{o} & \bfg{x}'-\bfg{x}_{o}' \end{bmatrix}^{\top}$.  Let $\bf A$ be the matrix that multiplies $\bfg{\zeta}$.  Then, the eigenvalues $\bar{\lambda}$ of $\bf A$ satisfy the generalized eigenvalue problem:
\begin{align}
    \label{eq:QEP}
    \det(\bar{\lambda}^2{\bf M}+\bar{\lambda}{\bf D}+{\bf K})=\det({Q}(\bar{\lambda}))=0.
\end{align}

The determinant can be turned into an ordinary polynomial, and its stability can be assessed accordingly, as follows:
\begin{align}
    \det({Q}(\bar{\lambda}))=a_4\bar{\lambda}^4+a_3\bar{\lambda}^3+a_2\bar{\lambda}^2+a_1\bar{\lambda}+a_0,
\end{align}
where the coefficients are: 
\begin{align*}
a_0 &= \det(\bf K) \, , \\  
a_1 &= \text{tr}(\text{adj}\,({\bf D} {\bf K})) \, , \\
a_2 &= \det(\bf D)+\text{tr}(\text{adj}\,({\bf M} {\bf K})) \, , \\
a_3 &= \text{tr}(\text{adj}\,({\bf M} {\bf D})) \, , \\
a_4 &= \det(\bf M) \, .
\end{align*}
The formulation of \eqref{eq:QEP} is in the form of \ac{qep}.  We refer the interested reader to \cite{Quadratic2001} for a discussion on this class of systems and a discussion of the properties of matrices $\bf{M}$, $\bf{D}$ and $\bf{K}$ related to system stability. 
  
In the context of the proposed \ac{ibr} control scheme, the generalized position is the vector $\bfg{u}_v = [\ln v, \theta_v]^{\top}$, where $v$ and $\theta_v$ are the magnitude and phase angle, respectively, of the terminal bus voltage of the converter.  Recalling the definition of complex frequency (see the Appendix), we observe that $\bfg \eta_v = \bfg{u}_v' = [\rho_v, \omega_v]^\top$ and, 
substituting into \eqref{eq_secord}, we obtain the second order system:
\begin{align}
\boxed{\bfg{\mathrm{M}}\bfg{\eta}_v'+\bfg{\mathrm{D}}\left(\bfg{\eta}_v-\bfg{\eta}_o\right)+\bfg{\mathrm{K}}\left(\bfg{u}_v-\bfg{u}_o\right)= \bfg{s}-\bfg{s}_o \, .}
\label{eq_main_gsc}
\end{align}
where $\bfg s = [p, q]^\top$ represents the input power vector for the \ac{ibr}, and $\bfg u_o$, $\bfg \eta_o$ and $\bfg s_o$ are set-point operating conditions.

Equation \eqref{eq_main_gsc} describes the power balance of each converter with the grid.  Then, the total energy of the \ac{ibr}, which is provided by the energy sources and storage that are exchanged with the grid, is as follows:
\begin{align}
    \Delta {E} &= \Re\left\{ \int_0^t{\left(\bfg{s}-\bfg{s}_o\right)\, dt} \right\} \\ \nonumber
    &= \Re \left\{ \int_0^t{\left(\bfg{\mathrm{M}}\bfg{\eta}_v'+\bfg{\mathrm{D}}\left(\bfg{\eta}_v-\bfg{\eta}_o\right)+\bfg{\mathrm{K}}\left(\bfg{u}_v-\bfg{u}_o\right)\right)\, dt} \right\} \, , 
\end{align}
where $\Delta {E}$ refers to the variation in the exchanged energy, 
\begin{equation}
\begin{aligned}
    \Re\left\{\int{\bfg{\mathrm{M}}\bfg{\eta}_v'\, dt}\right\}=&\int{\left(M_{11}\rho_v'+M_{12}\omega_v'\right)\, dt} \notag\\
    =&\ M_{11}\rho_v+M_{12}\omega_v+c_M \, 
\end{aligned}
\end{equation}
denotes the part of the exchanged energy that is provided by a virtual storage in variables $\rho_v$ and $\omega_v$,
\begin{equation}
\begin{aligned}
    \Re&\left\{\int{\bfg{\mathrm{D}}\left(\bfg{\eta}_v-\bfg{\eta}_o\right)\, dt}\right\}\notag\\
    &=\int{\left(D_{11}(\rho_v-\rho_o)+D_{12}(\omega_v-\omega_o)\right)\, dt}\notag\\
    &= D_{11}\left(\int{\frac{dv}{v}}-\int{\frac{dv_o}{v_o}}\right)+D_{12}\int{d(\theta-\theta_o)} \notag\\
    &= D_{11}\left(\ln{v}-\ln{v_o} \right)+D_{12}\left(\theta-\theta_o\right)+c_{\rm{D}} \, .
\end{aligned}
\end{equation}
the part of the exchanged energy that is provided by a virtual damping in $\ln{v}$ and the phase angle, and
\begin{equation}
\begin{aligned}
    \Re& \left\{\int{\bfg{\mathrm{K}}\left(\bfg{u}_v-\bfg{u}_o\right)\, dt}\right\}\notag\\
    &=\int{\left(K_{11}(\ln{v}-\ln{v_o})+K_{12}(\theta-\theta_o)\right)\, dt} \, ,
\end{aligned}    
\end{equation}
the part of the exchanged energy that is provided by the change in the potential energy.  We note that the system is rheonomous, that is, the integral is explicitly time dependent, and it cannot be calculated without knowing the trajectory of the voltage magnitude and its phase. 

Although in this work we define the control matrices as symmetric positive definite to satisfy theoretical properties, the practical implementation presents some challenges.  The aggregation of multiple \acp{gsc}, combined with the inherent nonlinearities of power system networks, makes it difficult to track the full matrices of the full system \ac{qep}.  As a consequence, extensive time-domain simulation analysis through Monte Carlo method are employed to improve the parameter tuning and assessment of the stability of \ac{gsc} implementations in the studied systems.

\subsection{Performance Metric}
\label{sec_performance}

To evaluate the dynamic performance of the proposed control, the index $\mu_{{\rm{ts}},k}$ is used to quantify variations in voltage magnitude and phase at bus $k$ taking into consideration transient and steady state.  The index is defined as a weighted sum of quadratic errors in comparison to initial and final steady-state operation points as follows:
\begin{equation}
  \label{eq.mu}
  \mu_{{\rm{ts}},k} = \mu_{{\rm{t}},k}+\mu_{{\rm{s}},k},
\end{equation}
where
\begin{align}
  \mu_{{\rm{t}},k} = \kappa_{{\rm{t}},\omega }\int(\omega_k-\omega_{{\rm{f}}})^2dt+\kappa_{{\rm{t}},\rho }
  \int (\rho_k-\rho_{\rm{f}} )^2 dt\,,\\
  \mu_{{\rm{s}},k} = \kappa_{{\rm{s}},\omega}(\omega_{\rm{o}}-\omega_{\rm{f}})^2+\kappa_{{\rm{s}},v }
 (v_{{\rm{o}},k}-v_{{\rm{f}},k} )^2 \,,
\end{align}
where the sub-indexes t and s stand for transient and steady-state behavior, and o and f refer to the initial and final conditions.  Thus, the parameters $\kappa_{{\rm{s}},\omega}$, $\kappa_{{\rm{s}},v}$, $\kappa_{{\rm{t}},\omega}$, $\kappa_{{\rm{t}},\rho}$ represent weights to assess the dynamic trajectory of the voltage phase and magnitude.  These parameters can be adjusted according to the requirements of each system and the control objectives.

The proposed metric quantifies the quadratic error of each \ac{cf} component with respect to its final steady state, while including a weighted measure of the difference in voltage magnitude and frequency between initial and final conditions.

To assess the overall dynamic impact of all devices on the system, we define the index $\mu_{\rm ts}$ as:
\begin{align}
\mu_{\rm ts} = \mu_{\rm t}+\mu_{\rm s} =\sum_{i \in \mathcal{G}}\left({\mu_{{\rm t},i}}+{\mu_{{\rm s},i}}\right),
\end{align}
where $\mathcal{G}$ represents the set of all buses in the grid in which a generator is connected.  This index provides a scalar measure of the total voltage variation across the devices of the system.  As $\mu_{\rm ts}$ is a cumulative metric, a smaller value indicates a more effective controller.

A similar performance index $\mu_k$ was defined in \cite{derfv} based on the concept of the total variations of the voltage at a node.  Nevertheless, the index in \cite{derfv} is unable to capture steady-state performance.

\section{Case Studies}
\label{sec_cases}

This section presents simulation results based on modified versions of the WSCC 9-bus system \cite{Sauer_Book} and the all-island Irish transmission system described in \cite{IRISH}.  In the first case, \acp{sm} are replaced by \acp{ibr} modeled using the proposed \ac{gsc} framework, according to the details provided in Section \ref{sec_control}.  For simplicity, all configuration settings are identical for all \acp{ibr} but are weighted by the generator base power.  For the Irish system, the behaviors of the different \ac{gsc} configurations are compared with the performance of an \ac{sm}-based system.

For cases where power unbalance is applied, load variation is assumed to be a \textit{sustained power perturbation}, namely, a step change of the grid power exchange that remains constant in the time scale of storage and control dynamics of the \ac{ibr}.  

To evaluate dynamic performance of the control, the index $\mu_{\rm{ts}}$ defined in Section \ref{sec_performance} is used to quantify variations in the voltage magnitude and phase at the generator buses within the transient and in steady state.

We explore the impact on stability of the proposed \ac{gsc} through a time-domain sensitivity analysis over the parameters of the \ac{gsc}.  This analysis is based on the Monte Carlo method, with a total of 20,000 realizations for each scenario.  Each realization consists on eigenvalue calculations of the linearized system and time-domain simulations, varying the studied parameters within an explicit range.  The sensitivity analysis for each parameter is used to calculate the performance metric $\mu_{\rm{ts}}$ and the unstable simulation rate.  
The latter is defined as the ratio between the number of unstable simulations and the total number of simulations conducted for a specific parameter range.   In particular, we assume that a simulation is stable if the following criteria are satisfied:
\begin{enumerate}
    \item The real and imaginary parts of the complex frequency of the internal voltage of each \acp{ibr} converge to a steady-state value within a specified tolerance; that is, $\bar{\eta}_k=\rho_{\rm{f}}+\jmath\, \omega_{\rm{f}}+\bar{\epsilon}$, where $\bar{\epsilon}$ is a given tolerance.
    \item The linearized system is asymptotically stable; that is, the real parts of the eigenvalues of the system are all negative: $\max(\Re\{\bar{\lambda}\})<0$.
    \item All modes of the linearized system are slower than a threshold given by the minimum acceptable time constant ${T}_{\min}$.  This is enforced by the following constraints:
    \begin{itemize}
        \item The decay rate of each mode must be slower than the threshold ${T}_{\min}$; particularly, the real parts must satisfy $-\Re\{\bar{\lambda}\}<{{T}}_{\min}^{-1}$.
        \item The oscillation frequency of each mode must be slower than the threshold ${T}_{\min}$; specifically, the imaginary parts must satisfy $|\Im\{\bar{\lambda}\}|<2\pi\,{T}_{\min}^{-1}$
    \end{itemize}
\end{enumerate} 

In this work, we adopt a tolerance $\bar{\epsilon}=10^{-10}(1+\jmath)$ and a minimum acceptable time constant ${T}_{\min}=0.01$ s.  As a conservative assumption, modes faster than those defined through ${T}_{\min}$ are considered unstable, as they can lead to instabilities due to possible couplings with very fast dynamics that are not considered in the model.

For all \acp{ibr}, the parameters of the LC filters are as follows: $L_{\rm f}=0.2$ mH , $R_{\rm f}=0$ $\rm \Omega$ and $C_{\rm f}=0.265$ $\rm \mu F$. Additionally, all weights for the index $\mu_{\rm ts}$ are considered with unitary magnitude.
All simulation results presented in this section were obtained using the simulation software tool Dome \cite{dome}.

\subsection{WSCC-9 bus system}

For this case study, we explore several configurations of the \ac{gsc}.  We begin by analyzing the \ac{vsm} and then extend it to a configuration with coupled power control by setting parameters such as $K_{12}$ and $M_{21}$ to non-zero values.  We also study the dual case of the \ac{vsm}, denoted as \ac{dvsm}, originally proposed in \cite{DualVSM2025}, where active and reactive power are still decoupled.  In this case, however, active power is linked to the real part of the \ac{cf} of the voltage $\rho_v$, and the reactive power is linked to its imaginary part, $\omega_v$.  While other configurations are possible, we focus on this set of examples to highlight the main features of the proposed controller.

\subsubsection{\ac{vsm} configuration}

The \ac{gsc} framework allows to choose a specific combination of parameters that makes the controller behave similarly to a \ac{vsm}.  This is achieved by setting parameters $M_{22}$, $D_{22}$, $D_{11}$, and $K_{11}$ non-null.  $M_{22}$ and $D_{22}$ represent the virtual inertia and damping of the frequency, respectively; and $D_{11}$ and $K_{11}$ act as a first order voltage droop controller.  With the rest of the parameters of the \ac{gsc} set as null, active and reactive power balance is mostly coupled with frequency and voltage variations, respectively.

Monte Carlo time-domain simulations are conducted ranging parameters $M_{22}$, $D_{22}$, $D_{11}$, and $K_{11}$ from 0.01 to 100.  A uniform logarithmic distribution is utilized for all parameters.  The contingency is a power unbalance given by a load variation of $\pm$0.5 pu in active and reactive power, where 0.5 pu corresponds to 15\% of the total active power load and 40\% of the reactive power load.  The sign of the load variation is randomized, while the magnitude is kept invariant.

\begin{figure}
    \centering
    \includegraphics[width=0.8\linewidth]{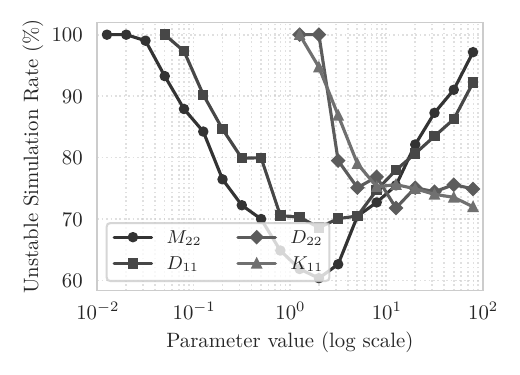}
    \caption{WSCC 9-bus system - Sensitivity analysis based on the Monte Carlo method for \ac{vsm} case - Unstable simulation rate as a function of parameter values in logarithm scale  under a load variation at bus 5 applied at second 1 of the simulations.}
    \label{fig:vsm_instability_rate}
    \vspace{-3mm}
\end{figure}

Figure \ref{fig:vsm_instability_rate} shows the unstable simulation rate in percentage for the WSCC 9-bus system and \ac{vsm} configuration.  The rate for each parameter ranges from 0.01 to 100 in logarithmic scale, and each point accounts for approximately 1000 simulations.
The unstable simulation rate in Fig.~\ref{fig:vsm_instability_rate} shows how each parameter impacts on the stability of the system.  $K_{11}$ stands for the stiffness in the variable $u_v=\ln{v}$, which has a similar effect on the stiffness in the voltage, and even more alike for voltages closer to the initial condition ($u_{\rm o}$).  It is interesting to note that for values smaller than 1, all simulations are unstable.  The same behavior is observed for $D_{22}$, which refers to the frequency damping controller.  On the other hand, parameters such as $M_{22}$ and $D_{11}$ are not allowed to be too small, as they are directly related with the time constants, and neither too big, as they do not allow to reach a steady state within the specific time simulation period ($t_{\rm f}=20$ s).

\begin{figure}
    \centering
    \includegraphics[width=0.8\linewidth]{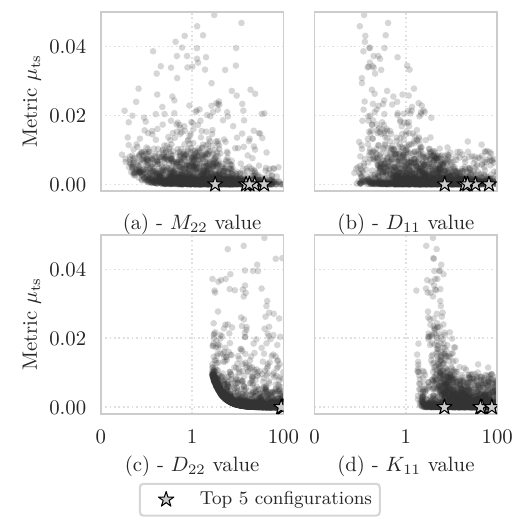}
    \caption{WSCC 9-bus system - Sensitivity analysis based on the Monte Carlo method for \ac{vsm} case - Performance metric $\mu_{\rm ts}$ as a function of parameter values in logarithm scale, $M_{22}$ in panel (a), $D_{11}$, in panel (b), $D_{22}$ in panel (c) and $K_{11}$ in panel (d).  Top 5 configurations are highlighted in each panel indicating the lowest metric values.}
    \label{fig:vsm_metric}
\end{figure}

Figure~\ref{fig:vsm_metric} shows the metric $\mu_{\mathrm{ts}}$ as a function of the parameters $M_{22}$ (panel a), $D_{11}$ (panel b), $D_{22}$ (panel c), and $K_{11}$ (panel d) on a logarithmic scale.  The top five configurations with the lowest metric values are highlighted in each panel.  A direct relation between $D_{22}$ and dynamic performance is observed, which is less evident for the other parameters.  

This can be explained by the construction of the $\mu_{\rm ts}$ metric, as it measures both the difference between the initial and final steady-state frequencies (which is clearly reduced by increasing $D_{22}$) and the deviation of the frequency trajectory from its final steady-state value.  The latter is minimized by a specific combination of parameters.  For instance, an overdamped system may take too long to reach steady state, increasing the transient metric, while a small $M_{22}$ can cause overshoot or oscillations, which also affects the metric.

\subsubsection{Extended \& Decoupled \ac{vsm} configuration}
\begin{figure}
    \centering
    \includegraphics[width=0.8\linewidth]{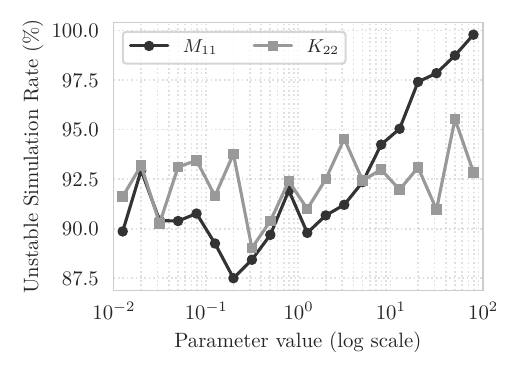}
    \caption{WSCC 9-bus system - Sensitivity analysis based on the Monte Carlo method  analysis for the extended \ac{vsm} case - Unstable simulation rate as a function of parameter values in logarithm scale  under a load variation at bus 5 applied at second 1 of the simulations.}
    \label{fig:vsm_e_instability_rate}
\end{figure}

\begin{figure}
    \centering
    \includegraphics[width=0.8\linewidth]{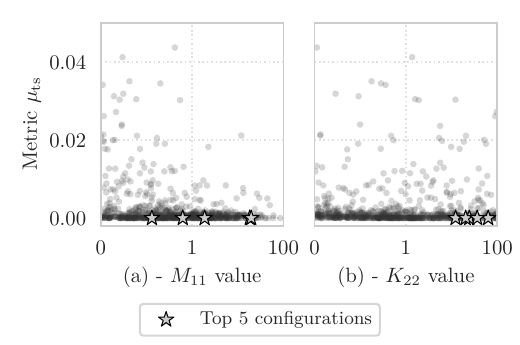}
    \caption{WSCC 9-bus system - Sensitivity analysis based on the Monte Carlo method for the extended \ac{vsm} case - Performance metric $\mu_{\rm ts}$ as a function of parameter values in logarithm scale, $M_{11}$ in panel (a) and $K_{22}$ in panel (b).  Top 5 configurations are highlighted in each panel indicating the lowest metric values.}
    \label{fig:vsm_e_metric}
    \vspace{-5mm}
\end{figure}

In this subsection the role of the parameters $M_{11}$ and $K_{22}$ is assessed.  Figure~\ref{fig:vsm_e_instability_rate} illustrates the unstable simulation rate as a function of these parameters in logarithm scale under a load variation at bus 5 applied at second 1 of the simulations.  Additionally, Fig.~\ref{fig:vsm_e_metric} displays the metric $\mu_{\mathrm{ts}}$ as a function of the parameters $M_{11}$ (panel a) and $K_{22}$ (panel b) on a logarithmic scale.

Figures \ref{fig:vsm_metric} and \ref{fig:vsm_e_metric} indicate that high values of $K_{22}$ tend to marginally reduce the metric and increase the unstable simulation rate.  Setting $K_{22}$ to a non-null and positive value introduces integral control for the frequency.  This can lead to several issues if not properly tuned, including unstable, unfair, or uneconomical active power distribution; windup problems; and poorly damped or even amplified oscillations.

\begin{figure}
    \centering
    \includegraphics[width=0.8\linewidth]{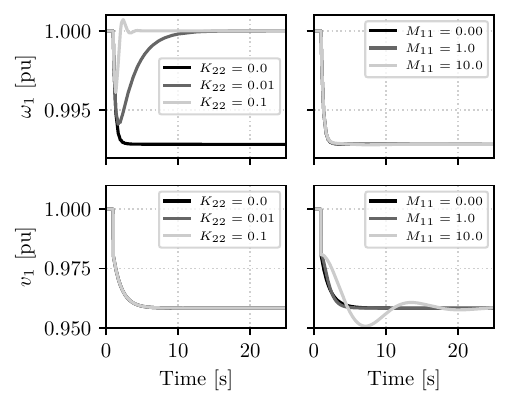}
    \caption{WSCC 9-bus system - Extended \& decoupled \ac{vsm} analysis - Effect on frequency and magnitude of the voltage behavior at bus 1, for a variation in parameters $K_{22}$ and $M_{11}$.}
    \label{fig:VSM_KM}
\end{figure}

The left panels of Fig.~\ref{fig:VSM_KM} illustrate the effect of $K_{22}$ in the frequency and voltage at the terminals of the \acp{ibr} for three parameter values: 0, 0.01, and 0.1.  The upper left panel shows how $K_{22}$ mitigates frequency error and that higher gain values imply faster controller response, but at the cost of increased oscillatory behavior.

Parameter $M_{11}$ acts as an inertia for the variable $u_v$, directly affecting the \ac{rocov}.  Although it does not have much influence in the metric for the studied range of the parameter as observed in Figure~\ref{fig:vsm_e_metric} (panel a), it has a significant effect on stability when increasing its value, as slower oscillatory responses do not reach a steady state in the simulation period.  
Finally, it is worth noting that voltage and frequency dynamics shown in Fig.~\ref{fig:VSM_KM} are not affected by $K_{22}$ and $M_{11}$, respectively, which indicates that these variables are dynamically decoupled.

\subsubsection{Coupled \ac{vsm} configuration}

In this subsection, a sensitivity analysis of the coupled terms in the active and reactive power balance is conducted.  To achieve this, we start from the \ac{vsm} and range terms such as $D_{12}$ and $D_{21}$ to couple the dynamic of the voltage magnitude with active power, and the dynamic of the voltage frequency to variations in reactive power.  These parameters are varied equally ($D_{12}=D_{21}$).  Thus, damping matrix $\mathbf{D}$ is still symmetric.  Symmetric damping matrices are often related to viscous damping, mutual impedance and proportional models, all of them of dissipative non-conservative nature.  On the other hand, skew-symmetric damping matrices are often associated with gyroscopic terms and Coriolis effects, being both conservative phenomena.

\begin{figure}[htbp]
    \centering
    \begin{minipage}{0.5\linewidth}
        \centering
        \includegraphics[width=\linewidth]{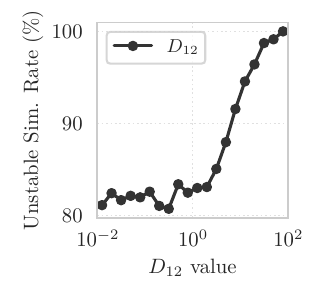}
        \subcaption*{(a)}
    \end{minipage}\hfill
    \begin{minipage}{0.5\linewidth}
        \centering
        \includegraphics[width=\linewidth]{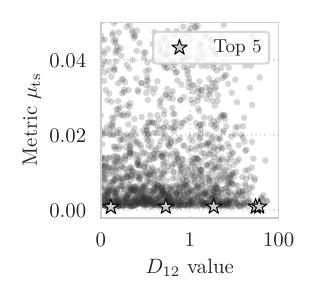}
        \subcaption*{(b)}
    \end{minipage}
    \caption{WSCC 9-bus system –  Coupled \ac{vsm} analysis.  
    Panel (a): Performance metric $\mu_{\rm ts}$ as a function of parameter $D_{12}$ in logarithm scale.  
    Panel (b): Unstable simulation rate as a function of $D_{12}$ values in logarithm scale under a load variation at bus 5 applied at second 1 of the simulations.  Top 5 configurations are highlighted.}
    \label{fig:vsm_c}
\end{figure}

Figure~\ref{fig:vsm_c} shows the unstable simulation rate (panel a), and the value of the performance metric $\mu_{\rm ts}$ (panel b) as a function of parameter $D_{12}$ ranging from 0.01 to 100.  Top 5 configurations are highlighted.

In transmission systems the $R/X$ ratio is relatively small, thus the inherent coupling between frequency and voltage with active and reactive power provided by the system becomes marginal.  This effect can be observed in the performance metric presented in Figure~\ref{fig:vsm_c}-(b), where the change of value for parameters $D_{12}$ and $D_{21}$ don't have a significant effect on $\mu_{\rm ts}$.  Nevertheless, increasing the value of the coupling parameter $D_{12}$ increases the unstable simulation rate, as it introduces non wanted interactions of local voltage variations into the frequency dynamics, and vice-versa.  

The effect of coupling parameters in the inertia and the stiffness matrix as well as the sensitivity in the R/X ratio is not studied in this paper and will be assessed in future work.  

\subsubsection{\ac{dvsm} configuration}
\label{sec_dvsm}

This subsection analyzes the \ac{dvsm} configuration. In this setup, the elements of the power vector $\bfg{s}=[p, q]^{\top}$ on the right-hand side of \eqref{eq_main_gsc} are interchanged, resulting in $\bfg{\tilde{s}} = [q, p]^{\top}$. The same transformation is applied to $\bfg{s}_o$. 

For the \ac{dvsm} configuration we set parameters $M_{22}$, $D_{22}$ $D_{11}$ and $K_{11}$ non-zero.  Thus, the \ac{dvsm} voltage dynamics is linked with active power and frequency dynamics with reactive power.
To address the main features of the \ac{dvsm}, a sensitivity analysis based on the Monte Carlo method is conducted for these parameters, ranging them between 0.01 to 100.

\begin{figure}
    \centering
    \includegraphics[width=0.8\linewidth]{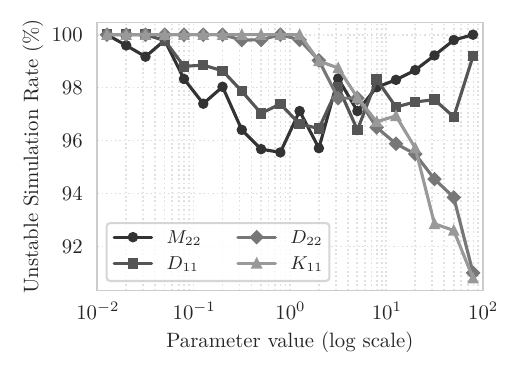}
    \caption{WSCC 9-bus system - Sensitivity analysis based on the Monte Carlo method  for \ac{dvsm} case - Unstable simulation rate as a function of parameter values in logarithm scale.}
    \label{fig:DVSM_instability_rate}
\end{figure}

Figure~\ref{fig:DVSM_instability_rate} illustrates the unstable simulation rate of the studied parameters.  It is interesting to note that in general the rate is higher, suggesting a narrower number of stable combinations.  Nevertheless, the shape of the rate as a function of the parameters is similar to the one displayed in Fig.~\ref{fig:vsm_instability_rate} for the \ac{vsm} configuration.

\begin{figure}
    \centering
    \includegraphics[width=0.8\linewidth]{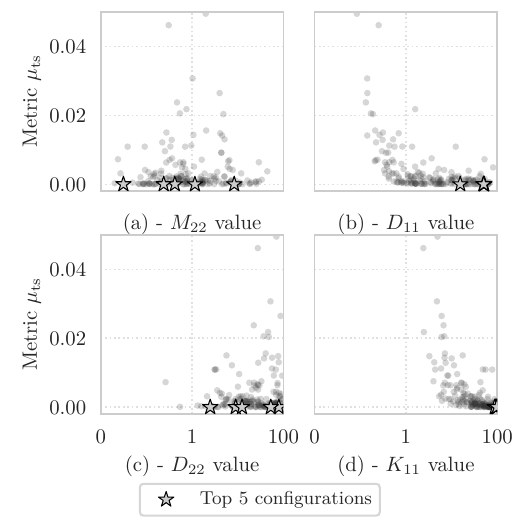}
    \caption{WSCC 9-bus system - Sensitivity analysis based on the Monte Carlo method for \ac{dvsm} case - Performance metric $\mu_{\rm ts}$ as a function of parameter values in logarithm scale, $M_{11}$ in panel (a), $M_{22}$ in panel (b), $D_{11}$ in panel (c), $D_{22}$ in panel (d), $K_{11}$ in panel (e) and $K_{22}$ in panel (f).  Top 10 configurations are highlighted in each panel indicating the lowest metric values.}
    \label{fig:DVSM_metric_1_1_1_1_}
\end{figure}

The performance metric assessment presented in Fig.~\ref{fig:DVSM_metric_1_1_1_1_} reveals a dual behavior among the studied parameters compared to the \ac{vsm} case.  This reflects that for the studied case, the active power balance has a major impact on the proposed metric, while reactive power control acts as a supporting tool for stability.  Furthermore, similar to the \ac{vsm} configuration, the results prioritize faster controllers, provided they do not couple with other time scales.

\subsection{Irish System}

The grid comprises 21 synchronous generators, 169 wind plants, and 2 HVDC interconnectors.  Additionally, there are 245 loads, 1479 buses, 796 lines, and 1055 transformers modeled within the system.  
Synchronous generators are modeled using Sauer and Pai's 6th order machine, simplified AVR IEEE type DC-1 with PSS2 and type 1 Turbine Governor, while wind turbines are modeled as \ac{dfig} (variable-speed 5th-order generator, double-mass elastic shaft with tower-shadow effect, turbine with continuous pitch control, cubic MPPT approximation, 1st-order AVR and turbine governor) and \ac{csig} (5th-order squirrel-cage induction generator model, turbine model without pitch control, single-mass shaft model with tower-shadow effect, and static capacitor bank) models.  While simulations are based on realistic data, they do not represent any specific operational condition of the system.

\begin{figure}[htb]
    \centering
    \includegraphics[width=0.99\linewidth]{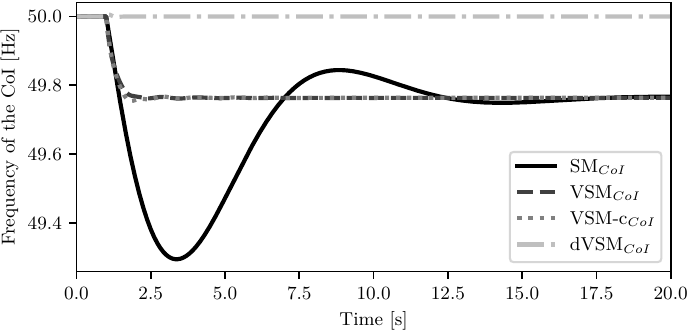}
    \caption{Irish system - Frequency of the \ac{coi} after the loss of the East to West Interconnector 500 MW for different     control setups of the \ac{gsc}.}
    \label{fig:Ire_w}
\end{figure}

\begin{figure}[htb]
    \centering
    \includegraphics[width=0.99\linewidth]{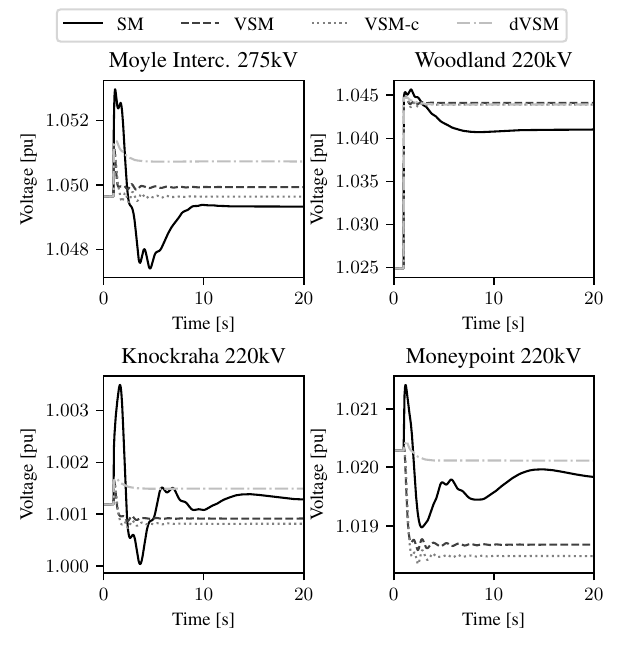}
    \caption{Irish system - Voltage at representative buses after the loss of the East to West Interconnector 500 MW for different
    control setups of the \ac{gsc}.}
    \label{fig:Ire_V}
\end{figure}

Figure \ref{fig:Ire_w} shows the frequency of the center of inertia for the Irish grid, whereas Fig.~\ref{fig:Ire_V} shows the voltage for various representative substations of the Irish grid located in the north, west, east and south of the system.  The contingency consists in the loss of the East to West interconnector, which is importing its nominal active power of 500 MW.  This is one of the most severe contingencies that can occur in the Irish transmission system in terms of power unbalance.  In this scenario, we assume that the \acp{sm} and \ac{ibr} have enough reserve to effectively provide the required active and reactive power lost during the event.

For the frequency of the \ac{coi} calculation, we use the definition of virtual inertia adopted in \cite{Tan2022} for systems with a significant share of \acp{ibr}, thus the frequency of the \ac{coi} is expressed as the following weighted sum:
\begin{align}
    \omega_{\rm CoI}= \frac{\sum_{k \in \mathcal{K}_{\rm SM}}H_{k}S_{b,k}\omega_k+\sum_{k \in \mathcal{K}_{\rm IBR}}H_{k}S_{b,k}\omega_k}{\sum_{k \in \mathcal{K}_{\rm SM}}H_{k}S_{b,k}+\sum_{k \in \mathcal{K}_{\rm IBR}}H_{k}S_{b,k}} ,
\end{align}
where $H_k$ are the physical or virtual inertia time constants and $S_{b,k}$ are the capacities of the generators.

For the case when \ac{gsc} is considered, all parameters for all devices are set equally.  Thus, the inertia constants are simplified, and the frequency of the \ac{coi} becomes a weighted sum with respect to the base power of each device.

We compare the dynamic performance across four scenarios.  The first is the base scenario without \ac{sm} replacement.  In the three remaining scenarios, \acp{sm} are replaced by \acp{ibr} with a \ac{gsc} scheme configured as \ac{vsm}; a \ac{vsm} with coupling parameters $D_{12}=D_{21}$ (denoted as VSC-c); and finally, a \ac{dvsm} scheme.  Parameters of $\bf{M}$, $\bf{D}$ and $\bf{K}$ are adjusted so that the initial \ac{rocof} and the steady-state value of the frequency and magnitude of the voltage are comparable.  For the particular case of the \ac{dvsm} we utilized the top values found in the Monte Carlo analysis within Section~\ref{sec_dvsm}.

Although all scenarios satisfy the Irish grid technical requirements for frequency and voltage during transient system disturbances ([48, 52] Hz and [200, 245] kV for 220 kV rated busbars \cite{eirgrid2025gridcode}), an overall improvement in the time to reach the steady state of the voltage across the buses of the system and the frequency of the \ac{coi} is observed.
The interaction among PSS and AVR of the \acp{sm} is clearly observed in the voltage transient through all the buses of the system.  This transient is mitigated due to the more homogeneous voltage and frequency control participation inherently provided by the \ac{gsc}.

Table~\ref{tab_metrics} presents the values of the metric described in Section \ref{sec_performance}.  This analysis evaluates the performance of three control strategies based on the \ac{gsc} (\ac{vsm}, \ac{vsm}-c, and \ac{dvsm}) following the loss of the 500 MW East-to-West Interconnector.  The \ac{vsm}-c configuration shows a slight improvement over the \ac{vsm} by exploiting the inherent trade-off and coupling between frequency and voltage dynamics.  

The \ac{dvsm} demonstrates a significant improvement in overall dynamic performance.  However, this performance gain is due to a fundamentally different power balancing mechanism compared to conventional strategies.  Due to the inductive nature of the transmission grid, the \ac{dvsm} has a local effect on the active power, which can sometimes be undesirable in systems with buses that have high voltage sensitivity and generators that lack significant power reserve.

\begin{table}[tb]
\centering
\caption{Irish system - Performance metric $\mu_{\rm ts}$ after the loss of the East to West Interconnector 500 MW for different control setups of the \ac{gsc}}
\label{tab_metrics}
\begin{tabular}{lccc}
\toprule
{Metric} & {VSM} & {VSM Coupled} & {dVSM} \\
\midrule
$\mu_{{\rm{t}},\rho}$ & \num{2.458e-05} & \num{4.006e-05} & \num{1.454e-04} \\
$\mu_{{\rm{t}},\omega}$ & \num{5.385e-04} & \num{5.365e-04} & \num{1.206e-07} \\
$\mu_{{\rm{t}}}$ & \num{5.631e-04} & \num{5.765e-04} & \num{1.455e-04} \\
\midrule
$\mu_{{\rm{s}},\omega}$ & \num{1.350e-02} & \num{1.350e-02} & \num{1.206e-07} \\
$\mu_{{\rm{s}},v}$ & \num{6.841e-03} & \num{6.372e-03} & \num{3.216e-03} \\
$\mu_{{\rm{s}}}$ & \num{2.034e-02} & \num{1.987e-02} & \num{3.216e-03} \\
\midrule
\textbf{$\mu_{{\rm{ts}}}$} & \num{2.090e-02} & \num{2.045e-02} & \num{3.362e-03} \\
\bottomrule
\end{tabular}
\end{table}

\begin{figure}[htb]
    \centering
    \includegraphics[width=0.99\linewidth]{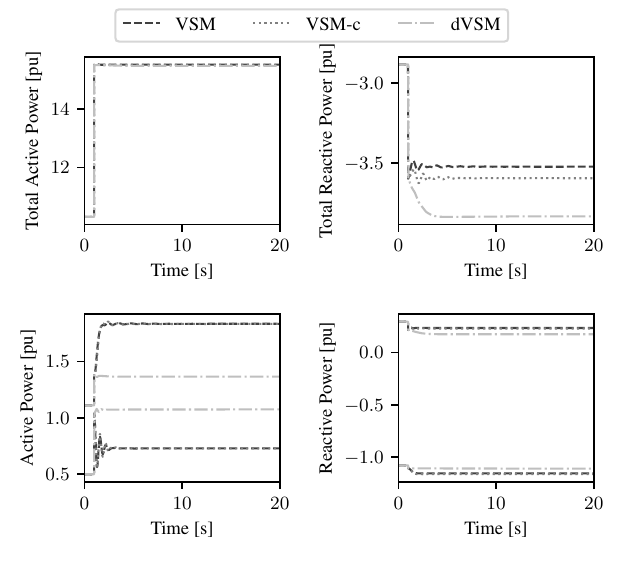}
    \caption{Irish system - Total active and reactive power (in pu, upper panels) provided by all \ac{gsc}-based generators, along with the active and reactive power provided by the generators located at buses \textsc{whitegat} and \textsc{pbegg4} (lower panels), following the loss of the 500 MW East-to-West Interconnector for different \ac{gsc} control setups.}
    \label{fig:Ire_pq}
\end{figure}

Figure~\ref{fig:Ire_pq} shows the total active and reactive power provided by all generators with different \ac{gsc} configurations in the upper panels, and the active and reactive power provided by the generators located at buses \textsc{whitegat} and \textsc{pbegg4} in the lower panels.
The total active power remains almost the same, but its distribution for the \ac{dvsm} configuration is different from that of the \ac{vsm} and the \ac{vsm}-c.  A similar effect occurs in the distribution of reactive power.  However, the total reactive power is significantly different across the three schemes.

\section{Conclusions}
\label{sec_conclusion}

This paper proposes a novel control framework, namely \acf{gsc}, which leverages the inherent flexibility of \acp{ibr} by extending the energy conversion principles of the swing equation to the complex domain.  The mathematical properties of the \ac{gsc} are analyzed within the \ac{qep} framework, with a focus on physical terms, namely inertia ($\bf{M}$), damping ($\bf{D}$), and stiffness ($\bf{K}$).

Monte Carlo simulations are employed to enhance parameter tuning and assess the stability of \ac{gsc} implementations.  Several \ac{gsc} configurations—including the \ac{vsm}, extended \ac{vsm}, coupled \ac{vsm}, and \ac{dvsm}—are evaluated using performance metrics such as the unstable simulation rate and the $\mu_{\rm ts}$ index (proposed to weight transient and steady-state variations).  Tests are carried out based on the WSCC 9-bus system and a realistic-size model of the all-island Irish transmission system.
Results demonstrate the framework's flexibility and its capability to enhance on power system stability.  The \ac{gsc} provides a unified approach to design \ac{gfm} schemes, effectively enhancing system damping and inertial response.

The proposed framework, currently designed for a two-dimensional vector space, can be extended to higher-dimensional spaces.  Therefore, future work will focus on utilizing the framework proposed in \cite{gfreq2022} to control the {\it geometric frequency} --- a generalization of the \ac{cf} and, consequently, of the conventional frequency.  Additionally, sophisticated implementations of the different configurations of \ac{gsc}, such as including current limiting, right-through modes, and parameter tuning, will be explored.

\appendix

The \ac{cf} is a physical quantity that can also act as a derivative operator of any complex number with non-null magnitude \cite{cmplx}.  For example, considering a complex time-dependent quantity, say $\bar{u}$, this can be written as:
\begin{equation}
  \bar{u} =
  u \exp(\jmath \, \alpha) =
  \exp(\ln{u}+\jmath \, \alpha) \, ,
\end{equation}
where $u\ne 0$ and $\alpha$ are the magnitude and phase angle of $\bar{u}$, respectively.  Assuming $\ln{u}$ and $\alpha$ are smooth functions of time, the time derivative of $\bar{u}$ gives:
\begin{equation}
  \begin{aligned}
    {\bar{u}'}
    &=\left ((\ln{u})' +
    \jmath \, {\alpha'} \right )\exp(\ln{u}+\jmath \, \alpha)\\
    &=\left(u'/u+\jmath \, \alpha' \right)\bar{u} 
    =\left(\rho_u +\jmath \, \omega_u \right)\bar{u} \\
    &=\bar{\eta}_u \, \bar{u} \, .
  \end{aligned}
  \label{eq.cmplx_freq}
\end{equation}

The quantity $\bar{\eta}_u$ is the \ac{cf} of $\bar{u}$, and $\rho_u$ and $\omega_u$ are its real and imaginary parts, respectively.


\begin{thebibliography}{10}
\providecommand{\url}[1]{#1}
\csname url@samestyle\endcsname
\providecommand{\newblock}{\relax}
\providecommand{\bibinfo}[2]{#2}
\providecommand{\BIBentrySTDinterwordspacing}{\spaceskip=0pt\relax}
\providecommand{\BIBentryALTinterwordstretchfactor}{4}
\providecommand{\BIBentryALTinterwordspacing}{\spaceskip=\fontdimen2\font plus
\BIBentryALTinterwordstretchfactor\fontdimen3\font minus
  \fontdimen4\font\relax}
\providecommand{\BIBforeignlanguage}[2]{{%
\expandafter\ifx\csname l@#1\endcsname\relax
\typeout{** WARNING: IEEEtran.bst: No hyphenation pattern has been}%
\typeout{** loaded for the language `#1'. Using the pattern for}%
\typeout{** the default language instead.}%
\else
\language=\csname l@#1\endcsname
\fi
#2}}
\providecommand{\BIBdecl}{\relax}
\BIBdecl

\bibitem{found}
F.~Milano, F.~D{\"{o}}rfler, G.~Hug, D.~J. Hill, and G.~Verbi{\v{c}},
  ``Foundations and challenges of low-inertia systems,'' in \emph{Power Systems
  Computation Conference (PSCC)}, 2018, pp. 1--25.

\bibitem{DynCoup2024}
H.~Cheng, C.~Li, A.~M. Y.~M. Ghias, and F.~Blaabjerg, ``Dynamic coupling
  mechanism analysis between voltage and frequency in virtual synchronous
  generator system,'' \emph{IEEE Trans. on Power Systems}, vol.~39, no.~1, pp.
  2365--2368, 2024.

\bibitem{ffr2025}
K.~Su \emph{et~al.}, ``Fast frequency response analysis for grid- following and
  grid-forming controlled bess considering voltage coupling effect,''
  \emph{IEEE Trans. on Power Delivery}, vol.~40, no.~4, pp. 2412--2425, 2025.

\bibitem{Laba2022}
Y.~Laba, A.~Bruyère, F.~Colas, and X.~Guillaud, ``{PQ} decoupling on
  grid-forming converter connected to a distribution network,'' in \emph{IEEE
  PEDG}, 2022, pp. 1--6.

\bibitem{derfv}
W.~Zhong, G.~Tzounas, and F.~Milano, ``Improving the power system dynamic
  response through a combined voltage-frequency control of distributed energy
  resources,'' \emph{IEEE Trans. on Power Systems}, vol.~37, no.~6, pp.
  4375--4384, 2022.

\bibitem{dercomp2024}
R.~Bernal and F.~Milano, ``Improving voltage and frequency control of ders
  through dynamic power compensation,'' \emph{Electric Power Systems Research},
  vol. 235, p. 110768, 2024.

\bibitem{cmplxctrl2025}
------, ``Complex frequency-based control for inverter-based resources,''
  \emph{Journal of Modern Power Systems and Clean Energy}, vol.~13, no.~5, pp.
  1630--1641, 2025.

\bibitem{TOULAROUD20232693}
M.~S. Toularoud, M.~K. Rudposhti, S.~Bagheri, and A.~H. Salemi, ``A
  hierarchical control approach to improve the voltage and frequency stability
  for hybrid microgrids-based distributed energy resources,'' \emph{Energy
  Reports}, vol.~10, pp. 2693--2709, 2023.

\bibitem{Wei2023}
W.~Du, K.~P. Schneider, G.~P. Wiegand, F.~K. Tuffner, J.~Xie, and O.~L. Dent,
  ``A supplemental control for dynamic voltage restorers to improve the primary
  frequency response of microgrids,'' \emph{IEEE Trans. on Smart Grid},
  vol.~14, no.~2, pp. 878--888, 2023.

\bibitem{VOC2022}
M.~Lu, ``Virtual oscillator grid-forming inverters: State of the art, modeling,
  and stability,'' \emph{IEEE Trans. on Power Electronics}, vol.~37, no.~10,
  pp. 11\,579--11\,591, 2022.

\bibitem{VSM_rev2022}
M.~Shadoul \emph{et~al.}, ``A comprehensive review on a virtual-synchronous
  generator: Topologies, control orders and techniques, energy storages, and
  applications,'' \emph{Energies}, vol.~15, no.~22, 2022.

\bibitem{VSClimit2020}
C.~Collados-Rodr\'iguez \emph{et~al.}, ``Stability analysis of systems with
  high vsc penetration: Where is the limit?'' \emph{IEEE Trans. on Power
  Delivery}, vol.~35, no.~4, pp. 2021--2031, 2020.

\bibitem{100gfl2020}
D.~Ramasubramanian and E.~Farantatos, ``Viability of synchronous reference
  frame phase locked loop inverter control in an all inverter grid,'' in
  \emph{IEEE PES General Meeting (PESGM)}, 2020, pp. 1--5.

\bibitem{bernal2025transientslackcapability}
\BIBentryALTinterwordspacing
R.~Bernal and F.~Milano, ``Transient slack capability,'' 2025. [Online].
  Available: \url{https://arxiv.org/abs/2505.17984}
\BIBentrySTDinterwordspacing

\bibitem{Trans2023}
Q.-H. Wu \emph{et~al.}, ``Transient stability analysis of large-scale power
  systems: A survey,'' \emph{CSEE Journal of Power and Energy Systems}, vol.~9,
  no.~4, pp. 1284--1300, 2023.

\bibitem{sun2018equal}
Y.~Sun, J.~Ma, J.~Kurths, and M.~Zhan, ``Equal-area criterion in power systems
  revisited,'' \emph{Proceedings of the Royal Society A: Mathematical, Physical
  and Engineering Sciences}, vol. 474, p. 20170733, 2018.

\bibitem{BoFan2022}
B.~Fan and X.~Wang, ``A {Lyapunov}-based nonlinear power control algorithm for
  grid-connected {VSCs},'' \emph{IEEE Trans. on Industrial Electronics},
  vol.~69, no.~3, pp. 2916--2926, 2022.

\bibitem{datadriven2021}
S.~Zhang, Z.~Zhu, and Y.~Li, ``A critical review of data-driven transient
  stability assessment of power systems: Principles, prospects and
  challenges,'' \emph{Energies}, vol.~14, no.~21, 2021.

\bibitem{DampInert2025}
C.~Feng, L.~Huang, X.~He, Y.~Wang, F.~Dorfler, and C.~Kang, ``Hybrid
  oscillation damping and inertia management for distributed energy
  resources,'' \emph{IEEE Trans. on Power Systems}, pp. 1--16, 2025.

\bibitem{GenInert2025}
Y.~Yu, Y.~Guan, W.~Kang, J.~C. Vasquez, and J.~M. Guerrero, ``A generic
  inertial-response preserved active-damping control for grid-forming inverters
  emulating synchronous machines,'' \emph{IEEE Trans. on Industrial
  Electronics}, vol.~72, no.~6, pp. 5897--5905, 2025.

\bibitem{GSE2022}
R.~Ma, J.~Li, J.~Kurths, S.~Cheng, and M.~Zhan, ``Generalized swing equation
  and transient synchronous stability with pll-based vsc,'' \emph{IEEE Trans.
  on Energy Conversion}, vol.~37, no.~2, pp. 1428--1441, 2022.

\bibitem{Quadratic2001}
F.~Tisseur and K.~Meerbergen, ``The quadratic eigenvalue problem,'' \emph{SIAM
  Review}, vol.~43, no.~2, pp. 235--286, 2001.

\bibitem{Sauer_Book}
P.~Sauer and M.~Pai, \emph{\BIBforeignlanguage{English (US)}{Power system
  dynamics and stability}}.\hskip 1em plus 0.5em minus 0.4em\relax Prentice
  Hall, 1998.

\bibitem{IRISH}
M.~Adeen \emph{et~al.}, ``On the calculation of the variance of algebraic
  variables in power system dynamic models with stochastic processes,''
  \emph{IEEE Trans. on Power Systems}, vol.~38, no.~2, pp. 1739--1742, 2023.

\bibitem{dome}
F.~Milano, ``A {Python}-based software tool for power system analysis,'' in
  \emph{IEEE PES General Meeting}, 2013, pp. 1--5.

\bibitem{DualVSM2025}
------, ``Dual grid-forming converter,'' \emph{IEEE Trans. on Power Systems},
  vol.~40, no.~2, pp. 1993--1996, 2025.

\bibitem{Tan2022}
B.~Tan \emph{et~al.}, ``Power system inertia estimation: Review of methods and
  the impacts of converter-interfaced generations,'' \emph{Int.~J.~of
  Electrical Power \& Energy Systems}, vol. 134, p. 107362, 2022.

\bibitem{eirgrid2025gridcode}
{EirGrid}, \emph{Grid Code Version {15}}, Jul.~23 2025, issue Date: 23 July
  2025, 463 pages.

\bibitem{gfreq2022}
F.~Milano, ``A geometrical interpretation of frequency,'' \emph{IEEE Trans. on
  Power Systems}, vol.~37, no.~1, pp. 816--819, 2022.

\bibitem{cmplx}
------, ``Complex frequency,'' \emph{IEEE Trans. on Power Systems}, vol.~37,
  no.~2, pp. 1230--1240, 2022.

\end{thebibliography}


\end{document}